\documentclass{article}

\usepackage{arxiv}

\usepackage[utf8]{inputenc} 
\usepackage[T1]{fontenc}    
\usepackage{url}            
\usepackage{booktabs}       
\usepackage{amsfonts}       
\usepackage{nicefrac}       
\usepackage{microtype}      
\usepackage{adjustbox}

\usepackage{xcolor}
\usepackage{colortbl}

\usepackage{caption}
\usepackage{amsfonts}
\usepackage{amsmath}
\usepackage{amssymb}

\usepackage{graphicx}
\usepackage[bottom]{footmisc} 
\usepackage{multirow}

\usepackage{floatrow}
\newfloatcommand{capbtabbox}{table}[][\FBwidth]

\title{RNNs on Monitoring Physical Activity Energy Expenditure in Older People}

\author{Stylianos~Paraschiakos
\thanks{{\tt\small email: s.paraschiakos@lumc.nl}}
\thanks{Leiden Institute of Advanced Computer Science, Universiteit Leiden, The Netherlands}
\thanks{Molecular Epidemiology, Dept. Biomedical Data Science, LUMC, Leiden, The Netherlands}
\and Cl{\'a}udio~Rebelo~de~S{\'a}\thanks{Data Science research group, University of Twente, Enschede, The Netherlands}
\and Jeremiah~Okai\footnotemark[2]
\and P. Eline Slagboom\footnotemark[3]
\and Marian~Beekman\footnotemark[3]
\and Arno~Knobbe\footnotemark[2]
}

\begin{document}
\maketitle

\begin{abstract}
Through the quantification of physical activity energy expenditure (PAEE), health care monitoring has the potential to stimulate vital and healthy ageing, inducing behavioural changes in older people and linking these to personal health gains. To be able to measure PAEE in a monitoring environment, methods from wearable accelerometers have been developed, however,  mainly targeted towards younger people. Since elderly subjects differ in energy requirements and range of physical activities, the current models may not be suitable for estimating PAEE among the elderly. Furthermore, currently available methods seem to be either simple but non-generalizable or require elaborate (manual) feature construction steps. Because past activities influence present PAEE, we propose a modeling approach known for its ability to model sequential data, the Recurrent Neural Network (RNN). To train the RNN for an elderly population, we used the Growing Old Together Validation (GOTOV) dataset with $34$ healthy participants of $60$ years and older (mean $65$ years old), performing $16$ different activities. We used accelerometers placed on wrist and ankle, and measurements of energy counts by means of indirect calorimetry. After optimization, we propose an architecture consisting of an RNN with $3$ GRU layers and a feedforward network combining both accelerometer and participant-level data. In this paper, we describe our efforts to go beyond the standard facilities of a GRU-based RNN, with the aim of achieving accuracy surpassing the state of the art. These efforts include switching aggregation function from mean to dispersion measures (SD, IQR, ...), combining temporal and static data (person-specific details such as age, weight, BMI) and adding symbolic activity data as predicted by a previously trained ML model. The resulting architecture manages to increase its performance by approximatelly $10\%$ while decreasing training input by a factor of $10$. It can thus be employed to investigate associations of PAEE with vitality parameters related to metabolic and cognitive health and mental well-being.

\end{abstract}


\section{Introduction}
\label{intro}
    
At older age, the extension of health span and maintenance of mobility are of great importance for the quality of life. Regular physical activity (PA) of moderate intensity is known to offer positive effects on the reduction of disease incidence and mortality risk~\cite{Manini2006, Chen2012, Cicero2012, Petersen2012}. To quantify and monitor the intensity of PA, estimation of energy expenditure during physical activity is an obvious necessity. By monitoring physical activity energy expenditure (PAEE), older people may better engage in physical activities, leading to better health and reduced mortality risk~\cite{Manini2006}.

PAEE is one component of total energy expenditure (TEE), where TEE is the sum of PAEE, resting energy expenditure (REE or RMR) by a fasted individual, and thermic effect of food (TEF). One way to measure PAEE is using direct calorimetry and measurements of heat production, but expensive equipment is required. Also, the Doubly Labeled Water Technique (DLW) provides an accurate technique of TEE estimation from where PAEE can be estimated, however, similar to direct calorimetry, it requires sophisticated lab-based equipment to analyze urine samples. Therefore, indirect calorimetry~\cite{Leonard2012} is commonly used, which  involves the measurement of oxygen and carbon dioxide exchange by ventilated mask or hood.

Because such forms of calorimetry cannot be performed under free-living conditions, methods to estimate PAEE from wearable accelerometers have been developed~\cite{Lyden2011,Staudenmayer2009, Ellis2014, Montoye2017, Caron2020, Driscoll2020}. This form of indirect calorimetry is estimated by accelerometer data and their combinations with physiological measurements such as heart rate, and individual-level data (demographic, anthropometric) using both linear and non-linear methods~\cite{Liu2012}. For example, linear or multiple regression methods can be used to estimate PAEE ~\cite{Lyden2011}, but also non-linear ensembles like random forest regressors~\cite{Gjoreski2013, Ellis2014, Driscoll2020} and deep learning method such as artificial neural networks (ANN)~\cite{Staudenmayer2009, Montoye2017} and convolutional neural networks (CNN)~\cite{Zhu2015} have been employed. Good estimates of PAEE can be derived from accelerometry data.

Since the majority of currently available methods to estimated PAEE from accelerometry data are mainly developed and tested on a young or middle-aged population~\cite{Montoye2017, Caron2020}, these models may not be suitable for estimating PAEE among the elderly. It is known that the elderly differ in energy requirements~\cite{Roberts2005, Horto2003}, expenditure~\cite{Frisard2007, Knaggs2011}, and range of physical activities~\cite{Lynette2009, Martin2014}.

There are two main drawbacks in the currently available methods: First, while linear models are pretty simple to deploy and use, they are unable to fit to all the activities~\cite{vanHees2009}. Second, the non-linear can be quite elaborate and computationally intensive, since they require steps of features construction and selection in order to capture the temporal nature of the accelerometer signal. Thus, a PAEE modeling method that does not require any sophisticated or hand-crafted pre-processing is called for, in addition to the development and testing of the model on older adults.

Therefore, we propose a neural network modeling approach that is known for its ability to model sequential data, the Recurrent Neural Network (RNN). The RNN is a network architecture that can deal with raw sensor data or minimum feature extraction, and can model temporal data by sequential processing. The nature of the processing in RNNs provides the possibility to remember information from the near as well as distant past, which is an advantage in comparison to ANN or CNN. Because past activities influence present PAEE, RNN modeling seems to be an excellent fit. 

To train the RNN for application on an elderly population, we used the Growing Old Together Validation (GOTOV) dataset~\cite{GOTOVpaper} with $34$ healthy participants of $60$ years and older (mean $65$ years old), performing $16$ different physical activities. This dataset is one of the first datasets publicly available with a focus on physical activity modeling of the elderly, both for activity recognition and energy expenditure. It includes multiple sensors (accelerometry, indirect calorimetry, physiological measurements) placed in multiple body locations. In the current study we used a combination of accelerometers placed on wrist and ankle (GENEActiv), because accelerometers combined on hand and foot can be good PAE estimators~\cite{Dong2013, Ellis2014}. Furthermore, Montoye \cite{Montoye2017,Montoye2017b} argues that both wrist and ankle separately produce the best PAEE estimations. Finally, the measurements of energy counts (per-breath calories) were collected by means of the medical-grade COSMED device~\cite{cosmed2001}.

Our proposed RNN architecture can make use of both accelerometer and participant-level data (age, gender, weight, height, BMI). This means that both temporal data and attribute-value data are given as input to the model and it combines them to give estimates of PAEE. In more detail, the model takes as an input sequences of temporal data representing a time window of past accelerometer, and creates output-features that are combined with the participant-level data in order to produce a PAEE estimation.

Summarizing, the main contribution of this paper is the development of a novel PAEE modeling architecture without any sophisticated feature construction step focused on a population group that is often overlooked: adults over $60$ years of age. The specific contributions of our work are the following:
\begin{enumerate}
    \item We proved that using statistical dispersion metrics (like standard deviation) to resample the accelerometer data to lower sampling rates can reduce the training time by approximately 10 times, compared to averaging.
    \item We model two different types of data, by taking advantage of both the temporal and the attribute-value nature of the data, accelerometer and participants-level data respectively.
    \item We proved that RRNs can estimate PAEE without prior knowledge of activity type.
\end{enumerate}

The rest of the paper is structured as following. Section~\ref{sec:data} presents the dataset used for model development. Then, Section~\ref{sec:method} discusses the methodological steps needed to model PAEE, such as model architecture, data preparation, model evaluation and experimental pipeline. This is followed by the results section (Section~\ref{sec:results}) presenting the main findings of our analysis. Finally, our findings, modeling strengths and limitations, and our future work is discussed in Section~\ref{sec:conclusion}.

\section{Dataset}
\label{sec:data}

The dataset used for our experiment is part of the Growing Old Together Validation (GOTOV) study~\cite{GOTOVpaper}. The GOTOV dataset is designed to develop both activity recognition~\cite{GOTOVpaper, EMBCpaper} and energy expenditure models that will serve multiple free-living ageing studies with similar population and devices \cite{GOTO, LLS, AGO}. The first part of the dataset, focused on activity recognition, is freely available in the $4$TU data repository\footnote{DOI: under publication, will be added here.  Access for reviewers please contact by email {\tt\small s.paraschiakos@lumc.nl}}.

One of the aims of this paper is to stimulate vitality-oriented research through the monitoring of physical activity among the elderly. For that reason, we extend the already public GOTOV data with data focused on energy expenditure. Thus, all calorimetry measurements combined with the ankle and wrist accelerometer data are made available from the $4$TU data repository\footnote{DOI: under publication, will be added here. Access for reviewers please contact by email {\tt\small s.paraschiakos@lumc.nl}}.

\subsection{Study Population}
GOTOV participants were recruited via newspaper advertisements and had to meet the following criteria:
\begin{enumerate}
\item Be older than 60 years old.
\item Have a healthy to overweight BMI\footnote{Body-Mass Index, the body mass divided by the square of the body height.} between $23$ and $35$ kg/m$^2$. 
\item Not being restricted in their movements by health conditions.
\item Bring their own bicycle.
\end{enumerate}

A total of $35$ individuals ($14$ female, $21$ male) between the ages $60$ and $85$ years old (mean $65$) and mean BMI $27$ kg/m$^2$ were recruited. The GOTOV study was approved by the Medical Ethical Committee of LUMC (CCMO reference NL38332.058.11). 

\subsection{Data Collection Protocol}
The $35$ participants performed a set of $16$ activities according to a specific protocol of approximately $90$ minutes. The $16$ activities were performed successively for specific time windows and with short breaks of standing still in between ($1$ minute). A researcher monitored the activities duration without giving any instructions or illustrations of the activities. The activity protocol took place at two locations; \textit{indoors} and \textit{outdoors} of the LUMC facility. The indoor activities consisted of \emph{lying down, sitting, standing, walking stairs} and several household activities, such as \emph{dish washing, staking shelves} and \emph{vacuum cleaning}. The indoor activities were performed in a room equipped with all the necessary instrumentation. The outdoor activities included different types of walking \emph{slow, normal, fast}, as well as \emph{cycling}. A visual example of the procedure can be found in a recorded video\footnote{\url{https://youtu.be/jvx5FGhqPxw}}.

Due to adverse weather conditions, only $25$ out of $35$ participants were able to perform the outdoor walking and cycling activities.

\subsection{Devices and body locations}

\begin{figure}[t]
    \centering
    \includegraphics[clip, trim= 0 0 0 0, width=1\textwidth]{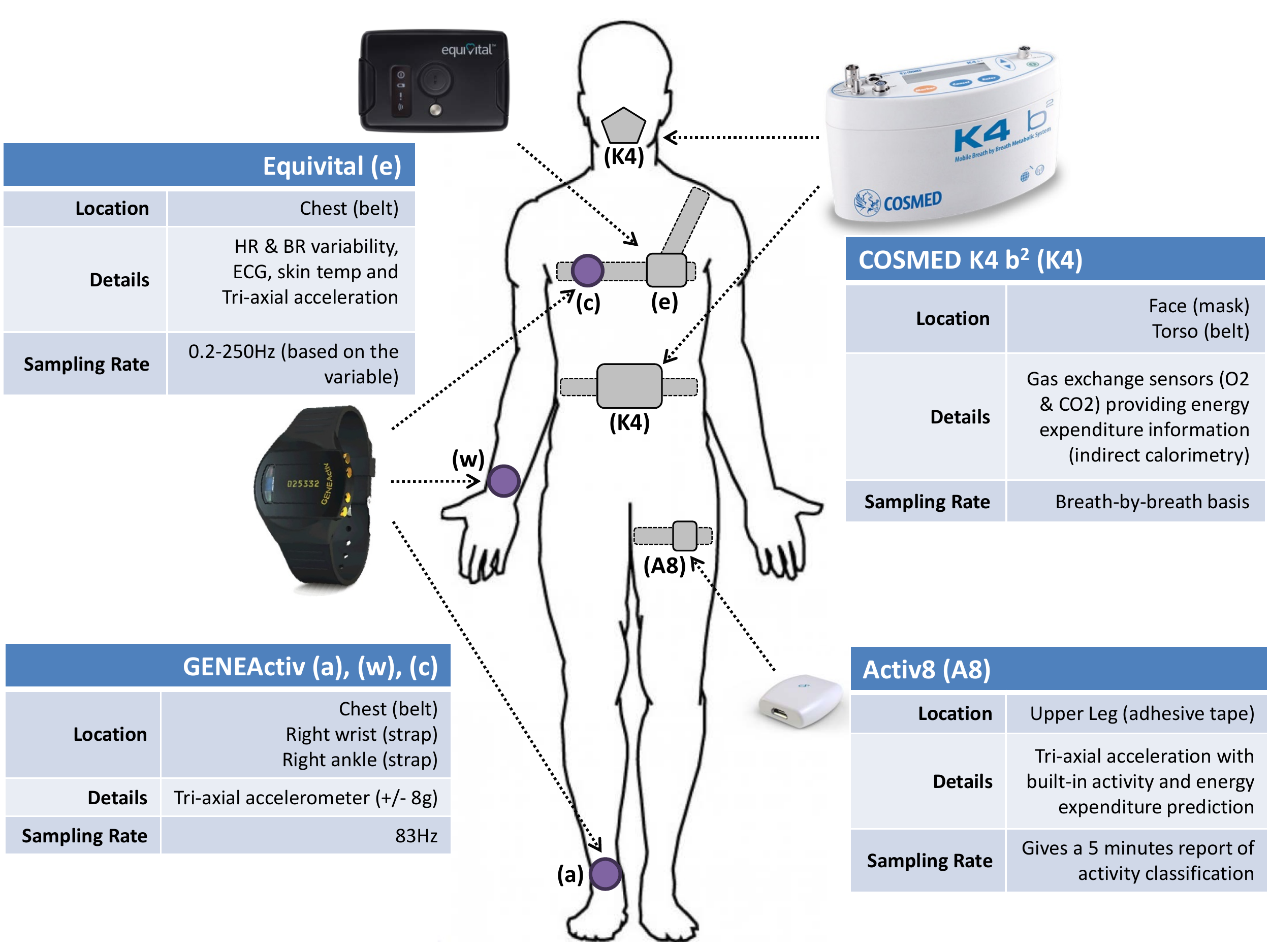}
    \caption{GOTOV study devices and their body location~\cite{GOTOVpaper}.}
    \label{fig:DevBody}
\end{figure}

During the data collection, the participant used $4$ different devices in $6$ body locations (see Figure~\ref{fig:DevBody}). The set of devices included both accelerometers and sensors measuring physiological indicators, e.g. indirect calorimetry (VO$_2$, VCO$_2$), breathing rate (BR) and heart rate (HR). In this study, we focus on the data coming from accelerometers and indirect calorimetry. This is mainly motivated by the fact that the models will serve existing free-living studies using the same sensor setup.

\paragraph{Accelerometry} The GENEActiv accelerometers placed on ankle wrist ({\bf a} and {\bf w} in Figure~\ref{fig:DevBody}) were used in order to recognise and measure activity levels of the participants. The GENEActiv accelerometers provided triaxial acceleration measurements ($\pm 8$ g) with a sampling rate of $83$ Hz. In order to create a recognisable pattern in data for synchronisation, the participants started the sequence of activities with a light jumping for $20$ seconds while waving arms.

\paragraph{Indirect calorimetry} The volume of oxygen (VO$_2$) and carbon dioxide (VCO$_2$) was measured per breath continuously during the activities, with a short break between the indoor and outdoor part of the protocol. The calorimetry measurements were obtained through the COSMED K4b$^2$~\cite{cosmed2001} device, with a portable unit on the torso and a flexible mask covering the participant's nose and mouth ($K4$ in Figure~\ref{fig:DevBody}). The mask is connected to the portable unit that contains O$_2$ and CO$_2$ analyzers, a sampling pump, a barometric sensors and electronics. The gas analyzer measures the exchange of oxygen and carbon oxygen (in ml\,kg$^{-1}$) and outputs PAEE metrics such as energy expended per minute, \emph{EEm} in Kcal per minute, or per hour, \emph{EEh} in Kcal per hour or \emph{METS}, where $1$ MET at rest $ = 1$ Kcal/kg/h. Measurements in these three units can be straightforwardly translated between one another. The COSMED metrics are calculated per breath based on formula that combines VO$_2$ and VCO$_2$ measurements and is similar\footnote{COSMED uses a slightly different estimation (unknown formula) giving an average overestimation of approx. 0.0098976 cal compared to Weir.} to the Weir formula~\cite{Weir1949}: 

\begin{center}
Metabolic rate (calories per minute) = $3.94\,VO_2+ 1.11\,VCO_2$ 
\end{center}

The output from this sensor in \emph{EEm} was used as our target for training and evaluating our PAEE estimation models.
The sampling rate (SR) of the target is equal to the breathing rate of the participant and depends also from the activity at a specific moment. This results in an SR that is not stable, with a mean SR among all existing data being equal to $0.3$ Hz.

The wearable unit weighs $1.5$ kg (battery included). Before every individual started the sequence of activities, the system was manually calibrated according to the manufacturer instructions. If the device was severely limiting a participant’s movement, it was removed and the participant was not involved in our analysis.

\subsection{Resulting Dataset}

\begin{table}[t]
\caption{Description of the final study population and their average COSMED measurements.}
\centering
\label{tab:partDetails}
\begin{tabular}{@{}l|rrr@{}}
\toprule
              & \begin{tabular}[c]{@{}c@{}}\textbf{Indoors} \\12 activities\end{tabular} & \begin{tabular}[c]{@{}c@{}}\textbf{Outdoors*}\\4 activities\end{tabular} & \begin{tabular}[c]{@{}c@{}}\textbf{Total}\\16 activities\end{tabular} \\ \midrule
\textbf{N female (\%)}              & $5~(27\%)$                                            & $6~(46\%)$                                                         & $11~(35\%)$     \\
\textbf{Age in years (SD)}          & $64.9~(4.4)$                                          & $66.8~(4.5)$                                                       & $65.7~(5.0)$    \\
\textbf{Height in cm (SD)}           & $176.2~(7.4)$                                         & $172.1~(8.3)$                                                      & $174.5~(7.9)$   \\
\textbf{Weight in kg (SD)}           & $83.7~(10.2)$                                         & $82.2~(13.5)$                                                      & $83.1~(11.5)$   \\
\textbf{BMI in kg/m$^2$ (SD)}         & $26.9~(2.0)$                                          & $27.7~(3.5)$                                                       & $27.2~(2.7)$    \\ \midrule
\textbf{EEm in Kcal (SD)}            &  $2.9~(0.4)$                                          & $6.1~(0.9)$                                                        & $3.8~(1.1)$     \\
\textbf{BR in breaths per sec (SD)}  & $0.30~(0.05)$                                         & $0.39~(0.04)$                                                      & $0.31~(0.04)$   \\\bottomrule
\end{tabular}
\\ \textbf{*}$4$ out of $18$ participants with outdoor activities did not perform cycling ($1$ female).
\end{table}

There were $35$ participants recruited in the GOTOV dataset, from whom $31$ participants had both COSMED (indirect calorimetry) and GENEActiv (ankle, wrist accelerometer) data. Of those, there were $13$ participants with only indoor activity data, so $12$ out $16$ activities. Finally, for all the other  participants with both indoor and outdoor activities, there were $4$ participants that did not perform the outdoor cycling activity.

Table~\ref{tab:partDetails} presents the participant-level data of this study and the average measurements of COSMED. In detail, in the first block it displays the number of female participants out of the total $31$ participants, and the average (mean and SD) age, height, weight and BMI. Furthermore, we can see the average EEm measurements by COSMED and breathing rate (sampling rate) for indoors, outdoors and total. From that, it is observed that there is a clear difference between the indoors and outdoors measurement in terms of EEm, where the mean outdoor EEm measurement is a bit more than double that of the indoor. This is something expected since the outdoor measurements include high intensity activities such as walking and cycling with a bigger range of EEm values compared to the indoors that have a smaller range. Similarly, the breathing rate is higher for the outdoor activities, which implies more data inputs for the same window of time when compared to the indoors (outdoors EEm SR higher than indoors), again as expected.

In total, the data set includes $2.8$ hours of sedentary activity (METs $< 1.5$), $5.4$ hours of light activity ($1.5\leq$ METs $< 4$), $1.8$ hours of moderate ($4\leq $ METs $< 6$) and 0$.73$ hours of vigorous activity ($6\leq$ METs).

\begin{figure}[t]
    \centering
    \includegraphics[clip, trim= 3cm 1cm 4cm 2.5cm, width=\textwidth]{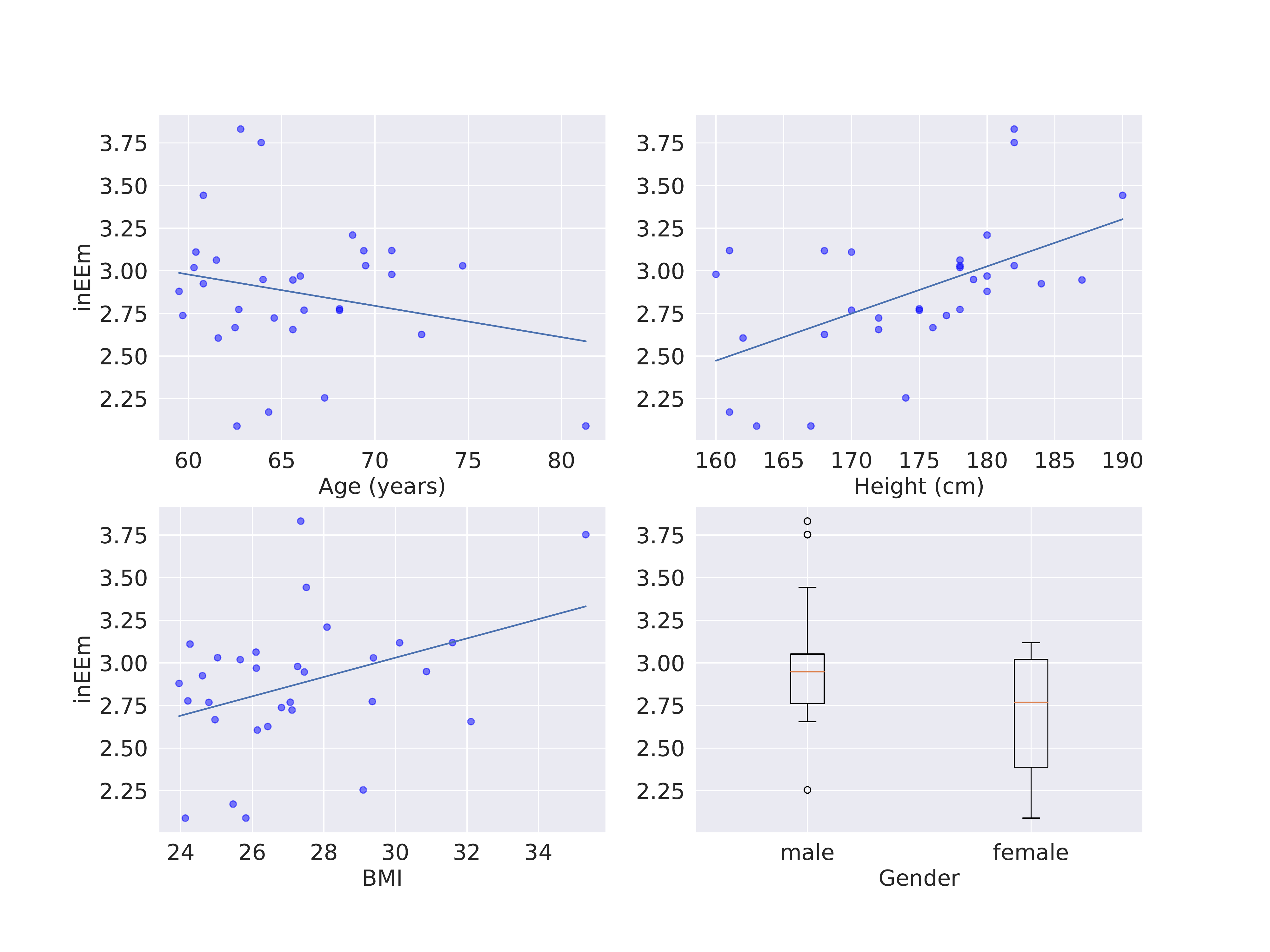}
    \caption{Trend of indoor activities Energy Expenditure across Age, Height, BMI and Gender.}
    \label{fig:dataset_inEEm}
\end{figure}

An initial view of the dataset is presented in Figure~\ref{fig:dataset_inEEm}, where the indoors energy expenditure measurements is plotted against gender, age, height and body composition per participant. We plotted the indoors EEm since all $31$ participants had indoors COSMED data. From the plots we see that the trends from the GOTOV dataset confirm what is known from the literature. In detail: 
\begin{itemize}
    \item EE decreases with age~\cite{Frisard2007, Roberts2005}. 
    \item EE increases with height~\cite{ARHills2014}.
    \item EE increases with body composition (BMI)~\cite{Weinsier1992}.
    \item EE in males is on average higher compared to the female participants~\cite{Keys1973}.
\end{itemize}

\section{Methodology}
\label{sec:method}

In this section, we explain the methodological contributions of the paper. In detail, we describe our model choice and its architecture. Following that, we analyse the steps of data preparation and their different combinations. Then, the training and evaluation process is explained. Finally, we summarize the experimental setup. 

\subsection{Modeling Architecture}
\label{sec:modeling}
A Recurrent neural network (RNN), is a type of artificial neural network that has the ability to `remember' older information from sequences. In more detail, an RNN contains feedback loops within its hidden layers whose activation at each time depends on that of the previous layer~\cite{GRU}. Consequently, RNNs have a modeling advantage when used on sequential or temporal data over traditional ANNs. RNNs have been used for a variety of tasks, such as natural language processing~\cite{LiX18}, speech recognition~\cite{LeePKL18}, and more recently, activity recognition from accelerometer data~\cite{Edel2016, Guan2017, EMBCpaper} and modeling of long-term human activeness~\cite{Kim2017}. Because PAEE is influenced by past activities (lag effect), RNNs could be suitable candidates for tackling the challenge of PAEE estimation.

Traditional RNN networks are known to struggle with information with long sequences due to the so-called \emph{vanishing gradient problem}~\cite{Hochreiter1998}. The most popular solution to this problem is introducing \emph{Long Short Term Memory} (LSTM) or Gated recurrent unit (GRU) layers. LSTM and GRU layers contain cells that act either as memory or gates controlling the information flow to the next layers. LSTMs contain $3$ gates, namely, \emph{forget, input} and \emph{output}, while GRUs have only $2$ gates, the \emph{reset} and \emph{update} gate. The reset gate controls which of the memory cell information needs to be forgotten. The update gate controls which information needs to be updated. This allows them to remember long sequences of information without losing relevant information. 

\begin{figure}[t]
    \centering
    \includegraphics[clip, trim= 0 0 0 0, width=\textwidth]{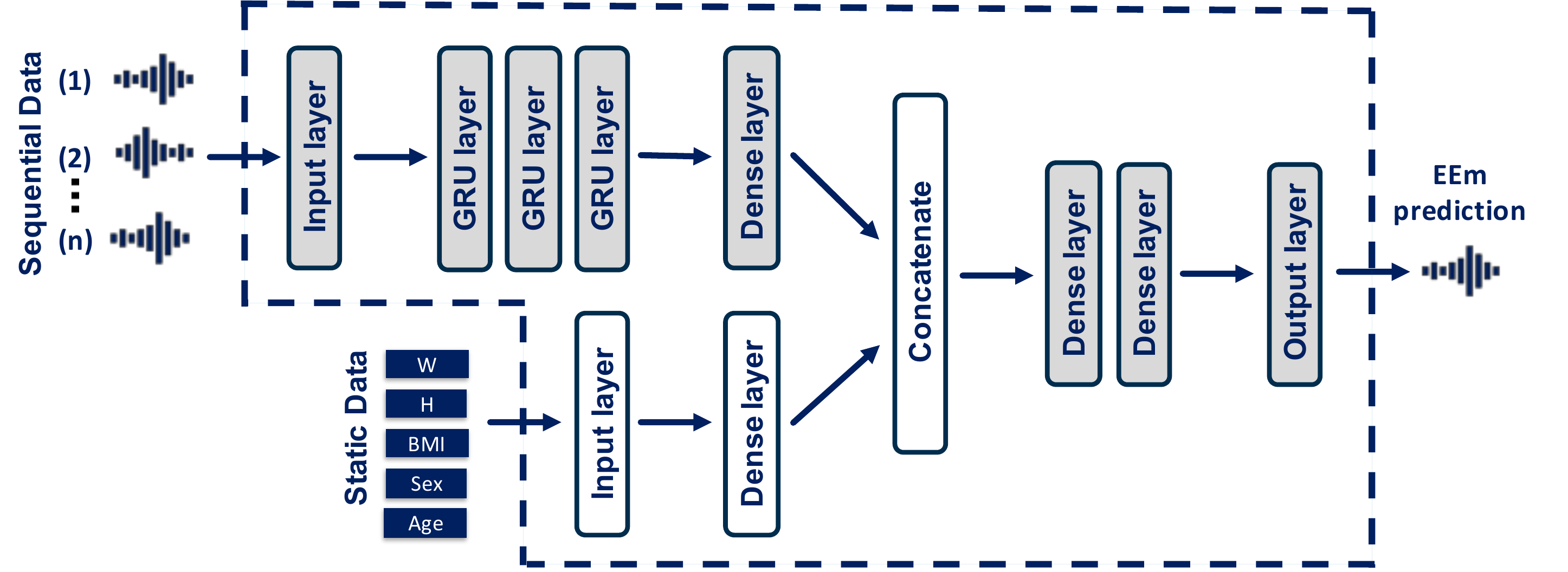}
    \caption{Proposed model architecture combining both time series and static data. The grey layers are sufficient when only temporal data is used (no static data).}
    \label{fig:modelArch}
\end{figure}

Our proposed RNN architecture consists of an input layer followed by $3$ GRU layers with $32$, $256$ and $32$ nodes respectively, $2$ dense layers with $32$ and $16$ nodes and an output layer (see Figure~\ref{fig:modelArch} grey layers). Models are trained to minimize the mean squared error (MSE) using a recently proposed optimization method called Adam~\cite{kingma2014adam}. Additionally, to prevent over-fitting, a dropout ratio of $0.5$ ($50\%$) is applied to all three GRU layers. We selected GRU cells over LSTM, as in our initial testing, they converged faster than LSTM while still preserving performance. 

Additionally, in order to test if participant-level data could improve PAEE estimation, we concatenated the aforementioned RNN setup and a single feedforward network, into a final feed-forward network demonstrated in Figure~\ref{fig:modelArch}. The reason behind such an architecture is the need to model two types of data, time series (sensor measurements, activities) and static data (participant-level). Therefore, we will need to feed the accelerometer sequences to the recurrent network with GRU layers and at the same time, we will need to feed the static data to a feedforward network. This feedforward network consists of an input layer and a hidden layer with $32$ neurons. The output layers of both networks were concatenated and connected to $2$ more hidden layers consisting of $32$, and $16$ neurons respectively. Finally, the output layer is made up of only a single neuron, which is used to predict the COSMED EEm values.

\subsection{Data preparation and choices}
\label{sec:prepro}
In order to build PAEE estimation models using RNNs, there is a need for several transformations both in the predictors data (accelerometers, activities, participant-level data) and the target (COSMED EEm). As a first step, target and numeric predictor data were $z$-normalized to have zero mean and a standard deviation of $1$. Additionally, in order to model discrete predictors, like gender or activity class, label encoding was used (with value between $0$ and $n$ classes-$1$).

\subsubsection{Indirect calorimetry as target data}
\label{sec:prepro:target}
COSMED produces energy expenditure metrics per breath, meaning that the target doesn't have a fixed sampling rate. On average, the COSMED sampling rate is $0.3$ Hz, which means one input approximately every $3$ seconds. In order to stabilize the sampling rate for the training, we resampled the COSMED signal to $0.1$ Hz by taking the mean of every interval of $10$ seconds. This way, we avoid the creation of more training data in periods with higher breathing rate, but we also smooth the outlier EEm values given by COSMED. Finally, we assign a sequence of predictors per EEm value which captures the movements that preceded the EEm measurement (see box $C$ of Figure~\ref{fig:dataPrepSteps}).

\subsubsection{Predictors data to sequences} 
\label{sec:prepro:pred}

To train the RNN model, we need to build sequences, where each is associated with one EEm measurement. A sequence is defined as a finite or infinite list of inputs arranged in a definite order~\cite{Volchan02}. For our problem, the order of inputs is based on time. The sequences represent the predictors data in the time immediately before every EEm measurement in windows of time, with specific number of inputs and resolution (sampling rate). 

\begin{figure}[t]
    \centering
    \includegraphics[clip, trim= 0 0 0 0, width=0.9\textwidth]{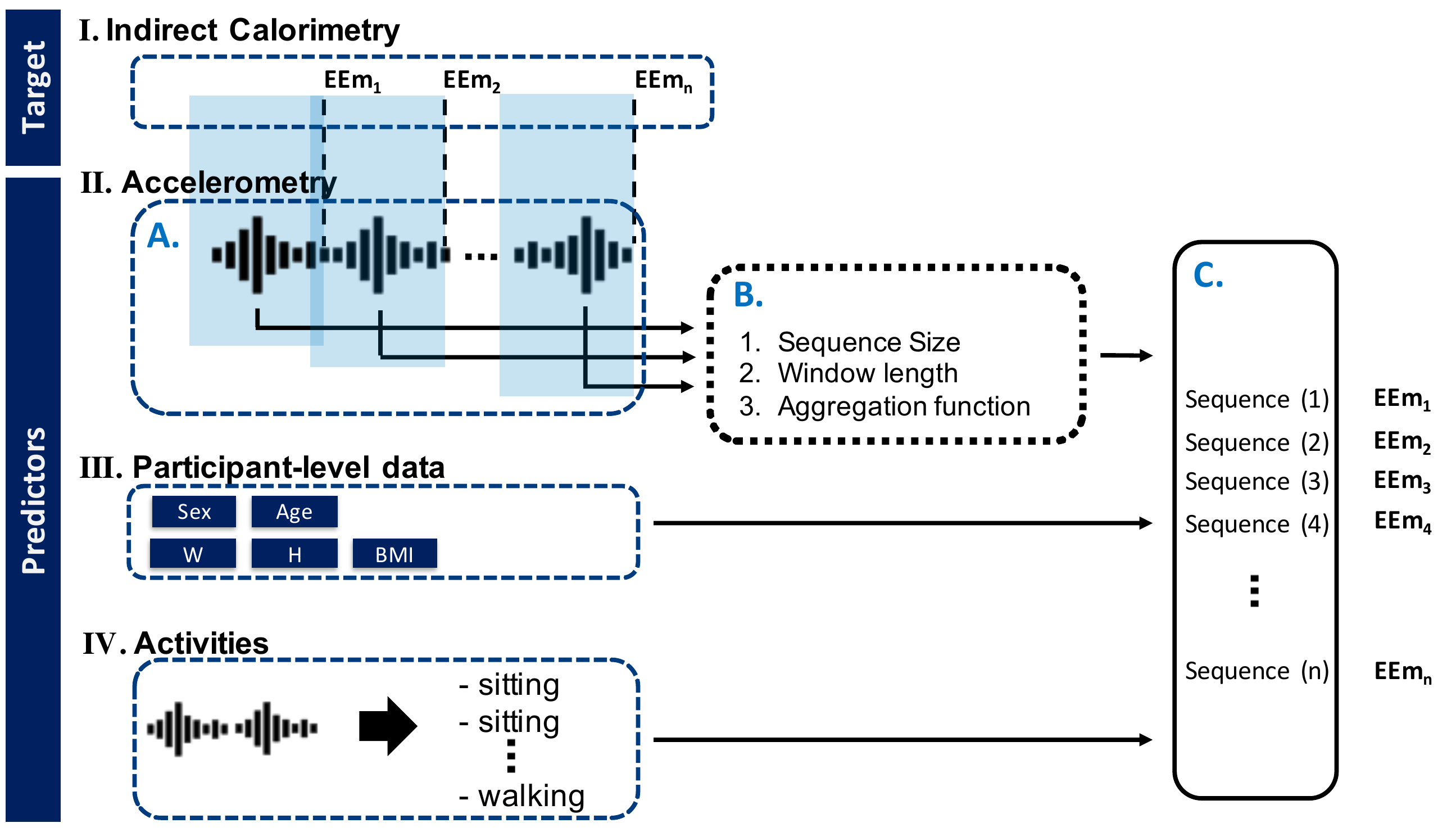}
    \caption{Building sequences for temporal data.}
    \label{fig:dataPrepSteps}
\end{figure}

We have two types of inputs in our network. First, the activity data is of a temporal nature, notably the accelerometer data (a numeric time series) and the activity labels (a discrete sequence). Second, the participant-level data that includes demographic information (age, gender) and body composition information (height, weight, BMI), as static attribute-value data. Figure~\ref{fig:dataPrepSteps} shows the different elements and steps taken to transform these data to training sequences. In detail, $I, II, III$, and $IV$ display the different inputs, while for the accelerometer data ($II$), we display the extra steps needed in order to transform the signal to training sequences.
In the following paragraphs, we explain the different sequence configurations developed and tested.

\paragraph{Accelerometers} 
In order to transform the accelerometer signal into training sequences, we need to decide on the number of inputs to be used (sequence size), the length of the window that those will represent (time interval) and the resolution (sampling rate). This is represented as the light blue shading in box \emph{A}, Figure~\ref{fig:dataPrepSteps}. For example, if we want a sequence with a size of $480$ inputs to represent a time window of $240$ s ($4$ minutes), we will need to resample the accelerometer data from $83$ Hz (original SR) to $2$ Hz, since the sampling rate (SR) depends on both sequence and window size.
\[
    SR = \frac{SequenceSize}{WindowSize},
\]
\noindent
where $WindowSize$ is calculated in seconds. 

For this reason, longer sequences can have higher resolutions of accelerometer data (high SR). However, they will be computationally more expensive to train. On the other hand, for a fixed sequence size, a choice of longer time windows will result in lower data resolution (low SR). Finally, the set of predictors' sequences will have the same number of sequences as the number of EEm values since every sequence is assiciated with one EEm input. We experimented with different sequence sizes representing different intervals of time (time windows) and data resolutions. The different decisions needed on resampling are displayed in box \emph{B} of Figure~\ref{fig:dataPrepSteps}.

In order to adjust the predictors to the given sequence sizes and window lengths, we need to resample the accelerometer data to the desired SR ($B4$, Figure~\ref{fig:dataPrepSteps}). We compared two different aggregation approaches, one that uses the mean function, and one that makes use of statistical dispersion functions (standard deviation, interquartile range, percentiles difference). In Section~\ref{sec:results}, we compare the performance of these two different approaches.

\paragraph{Participant-level data}
Combined with the accelerometer data, in this work, we test if participant-level data like demographics (age, gender) and anthropometric features (height, weight and BMI) could contribute to PAEE estimation ($IV$, Figure~\ref{fig:dataPrepSteps}). For this reason, we had to prepare such data input (standardization, one-hot encoding) and combine it in the accelerometer data sequences. This way the model will take as an input a sequence of accelerometer data and the details of its corresponding participant. 

\paragraph{Activity classes data}
Finally, we would like to test whether adding symbolic data in the form of a label describing the current activity can be beneficial to estimate PAEE ($III$, Figure~\ref{fig:dataPrepSteps}), when combined with either accelerometer data only, or with both accelerometer and participant-level data.
In order to obtain such activity labels, we had to first predict the activity types using learned activity recognition methods. We need to derive the labels from the acceleration data, since they wont we available in a free-living scenario either.

For this goal, we used a previously developed and published method that was already tested with the GOTOV devices~\cite{GOTOVpaper}. This model can produce activity predictions per second with an accuracy of more than $90\%$ based solely on ankle and wrist accelerometers for $7$-class activity classification. Through this model, we can predict the following $7$ classes: \emph{lying down, sitting, standing, household, walking, cycling} and \emph{jumping}. Having the activity labels predicted per second, we encoded them and combined them with the accelerometer sequences as input to our model. Table~\ref{tab:EEm_perAct} summarizes the predicted classes and presents also some statistics about their EEm cost.

\begin{table}[t]
\caption{Table of time spent and EEm (Kcal/minute) per activity.}
\label{tab:EEm_perAct}
\begin{tabular}{@{}l|rr@{}}
\toprule
\multicolumn{1}{c|}{\textbf{Activity}} & \textbf{Time (hours)} & \textbf{EEm (SD)} \\ \midrule
\textbf{lying down}                    & 0.8 (7.4\%)           & 2.4 (1.1)         \\
\textbf{sitting}                      & 1.6 (14.8\%)          & 2.0 (1.0)         \\
\textbf{standing}                     & 2.5 (23.2\%)          & 3.2 (1.7)         \\
\textbf{household}                    & 2.2 (20.4\%)          & 3.5 (1.6)         \\
\textbf{walking}                      & 2.3 (21.4\%)          & 5.2 (2.1)         \\
\textbf{cycling}                      & 1.3 (12.1\%)          & 8.2 (3.0)         \\
\textbf{jumping}                      & 0.1 (0.6\%)           & 3.1 (1.7)         \\
\bottomrule
\end{tabular}
\end{table}

\subsection{Training and Evaluation}
\label{sec:eval}

We trained and tested our models using \emph{Leave One Subject Out Cross-Validation} (LOSO-CV). This means that we train using all subjects (participants), leaving the data of one subject out as a test set. We then iterate the process in order to test all subjects separately. The aim of this type of cross-validation is that we emulate the future situation where we would like to process as yet unseen subjects. The LOSO-CV process prevents training set leakage within a subject, as normal cross-validation procedures might allow. Additionally, during training, $2$ participants were selected as validation set, one with only indoor activities and one with all activities. These $2$ sets were randomly chosen per subject and were the same across the different model settings tested in order to have fair comparisons. All models were trained for $50$ epochs with a batch size of $512$. 

We would also like to point out that the model's validation and test sets during LOSO are used with their original sampling rate (once per breath). This means that during validation and testing, we evaluate our models on the original COSMED data This means that we \emph{trained} our models using the smoothed EEm values with a stable SR ($0.1$ Hz), as indicated in the previous section, but we \emph{evaluate} them per breath. This way, we can see which model can fit better the input data since we evaluate our models by measuring their performance on the original EEm values, including the extreme COSMED measurements.

Hence, to get the overall performance of a model, we train $31$ different models using LOSO and we report the aggregated (median) result, as \emph{Root Mean Squared Error} (RMSE) and \emph{R-squared} (R$^2$). Similar to that, we compute the RMSE and R$^2$ separately for indoors and outdoors data. The reason behind this is their big differences in magnitude (see Table~\ref{tab:partDetails}). Additionally, since there are participants without outdoors data, we can see how our models behave on low and high-intensity activities.

\subsection{Experimental Pipeline}
\label{sec:expSetUp}

In this section, we explain the experiments performed and their order. First, we tested our proposed architecture with only accelerometer data (grey in Figure~\ref{fig:modelArch}) comparing the different accelerometer aggregation functions into time windows of different sequence size and resolution (SR). The aggregation functions tested were \emph{mean, standard deviation (SD), interquartile range (IQR)}, and \emph{difference between $5$th} and \emph{$95$th percentile (PD)}, with:
\begin{itemize}
    \item sequence sizes of $4, 10, 50, 160, 240, 360$, and $480$ inputs per sequence, and
    \item window lengths, for each sequence size, of  $1,2,4$, and $8$ minutes.
\end{itemize}
Every combination of sequence size, window lengths, and aggregation function were tested, concluding to different resolutions (SR) per window. In total, $28$ different combinations were tested.

As a second analysis step, we test whether either the addition of participants-level data or the activity classes improves the performance. In order to do that, we make use of the complete architecture presented in Figure~\ref{fig:modelArch} and test the different combinations of additions. In detail, we compare the performance of the model using: $1)$ both accelerometer and participants level data (GA\_ID); and $2)$ using accelerometer, participant-level and activity classes data (GA\_ID\_AC). 

Subsequently, in order to have comparable results with literature, we compare our selected model's performance over different EEm aggregations and activity. Other than that, in a free-living setup, where breathing rate information is not included, PAEE would be estimated per specific time windows. In particular, we report the performance across different EEm windows from original COSMED SR (breath by breath), to $10$, $30$, $60$ seconds and $5$, $60$ minutes aggregations and per activity. This way, we can have an overview of how our approach can be used to estimate PAEE of longer windows.

\section{Results}
\label{sec:results}

\subsection{Standard deviation as optimal aggregation function}

In Table~\ref{tab:resultsTab}, the performance of the different input data setups is presented. Here, we present the best setup per aggregation function since it is not feasible to display all $28$ combinations. In the first column the aggregation function is displayed, followed by its concluding best data setup (sequence size, window length, sampling rate distribution). Then we compare their R$^2$ and RMSE in total, indoor, and outdoor activities. Additionally, in last column we present if there is any significant difference (with p$<0.05$) between mean and the rest functions in terms of R$^2$, using a paired t-test. 

\begin{table}[t]
\caption{Table comparing the performance of different data setups. The p-value corresponds to the paired t-test of mean with SD, IQR, in terms of R$^2$, with \textbf{*} pointing out when p$<0.05$.}
\label{tab:resultsTab}
\centering
\begin{tabular}{@{}ll|
>{\columncolor[HTML]{EFEFEF}}c c
>{\columncolor[HTML]{EFEFEF}}c |c
>{\columncolor[HTML]{EFEFEF}}c c|c@{}}
\toprule
\multicolumn{2}{c}{\textbf{Model}} & {\color[HTML]{000000} \textbf{R$^2$}} & {\color[HTML]{000000} \textbf{inR$^2$}} & {\color[HTML]{000000} \textbf{outR$^2$}} & \textbf{RMSE} & \textbf{inRMSE} & \textbf{outRMSE} & \textbf{p-value} \\ 
\midrule
                & SeqSize = $480$     &                                       &                                         &                                          &               &                 &                  &                \\
              & WinSize = $4$min & $0.38$                                & $0.23$                                  & $0.38$                                   & $1.46$        & $1.24$          & $2.32$           & -           \\
\multirow{-3}{*}{\textbf{mean}}  & SR = $2$ Hz                    &                                         &                                          &               &                 &                  &                \\
\midrule
              & SeqSize = $50$     &                                        &                                         &                                          &               &                 &                  &                \\
              & WinSize = $2$ min & $0.45$                                & $0.31$                                  & $0.33$                                   & $1.35$        & $1.16$          & $2.03$           & $\mathbf{p<0.01}$\textbf{*}          \\
\multirow{-3}{*}{\textbf{SD}}  & SR = $0.42$ Hz                     &                                         &                                          &               &                 &                  &                \\ 
\midrule
              & SeqSize = $50$     &                                       &                                         &                                          &               &                 &                  &                \\
              & WinSize = $2$ min & $0.41$                                & $0.29$                                  & $0.33$                                   & $1.39$        & $1.18$          & $2.15$           & \textbf{0.02*}           \\
\multirow{-3}{*}{\textbf{IQR}} & SR = $0.42$ Hz                      &                                         &                                          &               &                 &                  &                \\ 
\midrule
              & SeqSize = $50$     &                                       &                                         &                                          &               &                 &                  &                \\
              & WinSize = $2$ min & $0.43$                                & $0.30$                                  & $0.28$                                   & $1.36$        & $1.17$          & $2.03$           & $\mathbf{p<0.01}$\textbf{*}          \\
\multirow{-3}{*}{\textbf{PD}} & SR = $0.42$ Hz                      &                                         &                                          &               &                 &                  &                \\ 
\bottomrule
\end{tabular}
\end{table}

Here, we observe that models built with statistical dispersion functions outperform significantly the one using the mean, for $\alpha=0.05$ all p's are $<\alpha$. Nevertheless, the mean-based model even with approximately $10$ times more data inputs (sequences of $480$ versus $50$ inputs) and double window size ($4$ versus $2$ minutes) it does not achieve a similar performance. Furthermore, comparing only the models built with SD, IQR and PD, there is no clear performance difference. That is because all these three measures are really similar and their main differences are in the magnitude of training values. Nevertheless, if we have to choose one of them, we believe that the SD model seems to be slightly better than the others, both in terms of R$^2$ and RMSE. Adding to that, standard deviation is more intuitive as a metric compared to IQR and PD. Therefore, for the rest of our analysis, we will focus on the model built with the following settings for accelerometer data: $1)$ a sequence size of $50$ inputs, $2)$ representing a time window of $2$ minutes, $3)$ resampled to a resolution of $SR=0.42$ Hz with SD. 

\subsection{Adding participants-level results to better PAEE estimations}
Now we have an RNN model using accelerometer data resampled by SD in a window of $2$ minutes to estimate PAEE (GA, in Table~\ref{tab:ga_id_ac}), we investigated whether addition of participant-level data would improve the estimations. The use of participant level data (such us age, sex, height, weight, and BMI) besides accelerometer data, improves the results of EEm estimation, both in terms of R$^2$ and RMSE error (GA\_ID model). In more detail, models' performance improves significantly ($p-value=0.02$) from $0.45$ (GA) to $0.55$ (GA\_ID) for R$^2$ while for RMSE from $1.35$ Kcal/min the error decreased to $1.25$ (Table~\ref{tab:ga_id_ac}).

\begin{table}[t]
\centering
\caption{Comparing models with participant-level data and activity classes. The p-value corresponds to the paird t-test of GA (only accelerometry data) with any extra data addition, GA\_ID, GA\_AC, and GA\_ID\_AC, in terms of R$^2$, with \textbf{*} pointing out when p$<0.05$.}
\label{tab:ga_id_ac}
\begin{tabular}{@{}r|
>{\columncolor[HTML]{EFEFEF}}c c
>{\columncolor[HTML]{EFEFEF}}c| c
>{\columncolor[HTML]{EFEFEF}}c c|c@{}}
\toprule
           & \textbf{R$^2$}  & \textbf{inR$^2$}   & \textbf{outR$^2$}  & \textbf{RMSE}   & \textbf{inRMSE}  & \textbf{outRMSE} & \textbf{p-value} \\ \midrule
\textbf{GA}         & $0.45$ & $0.31$    & $0.33$    & $1.35$ & $1.16$  & $2.03$  & -                 \\
\textbf{GA\_ID}     & $0.55$ & $0.37$    & $0.36$    & $1.25$ & $1.09$  & $2.05$  & $\mathbf{0.02}$\textbf{*}   \\
\textbf{GA\_AC}     & $0.42$ & $0.32$    & $0.35$    & $1.33$ & $1.14$  & $1.96$  & $0.38$            \\
\textbf{GA\_ID\_AC} & $0.50$ & $0.33$    & $0.40$    & $1.29$ & $1.13$  & $2.00$  & $0.30$            \\ 
\bottomrule
\end{tabular}
\end{table}

\subsection{Activity classes do not contribute to PAEE estimations}
Interestingly, adding activity labels doesn't seem to improve the results (see GA\_AC in Table~\ref{tab:ga_id_ac}), where the R$^2$ drops, not significantly though, and RMSE increases. When combining both participant-level data and activity classes (GA\_ID\_AC) it seems that there is a slight improvement (R$^2$=$0.50$) in the results but not significantly ($p=0.3>\alpha$). 

\subsection{Demonstration of the RNN model estimating}
\begin{figure}[t]
    \centering
    \includegraphics[clip, trim= 3.5cm 1cm 3.5cm 2.3cm, width=\textwidth]{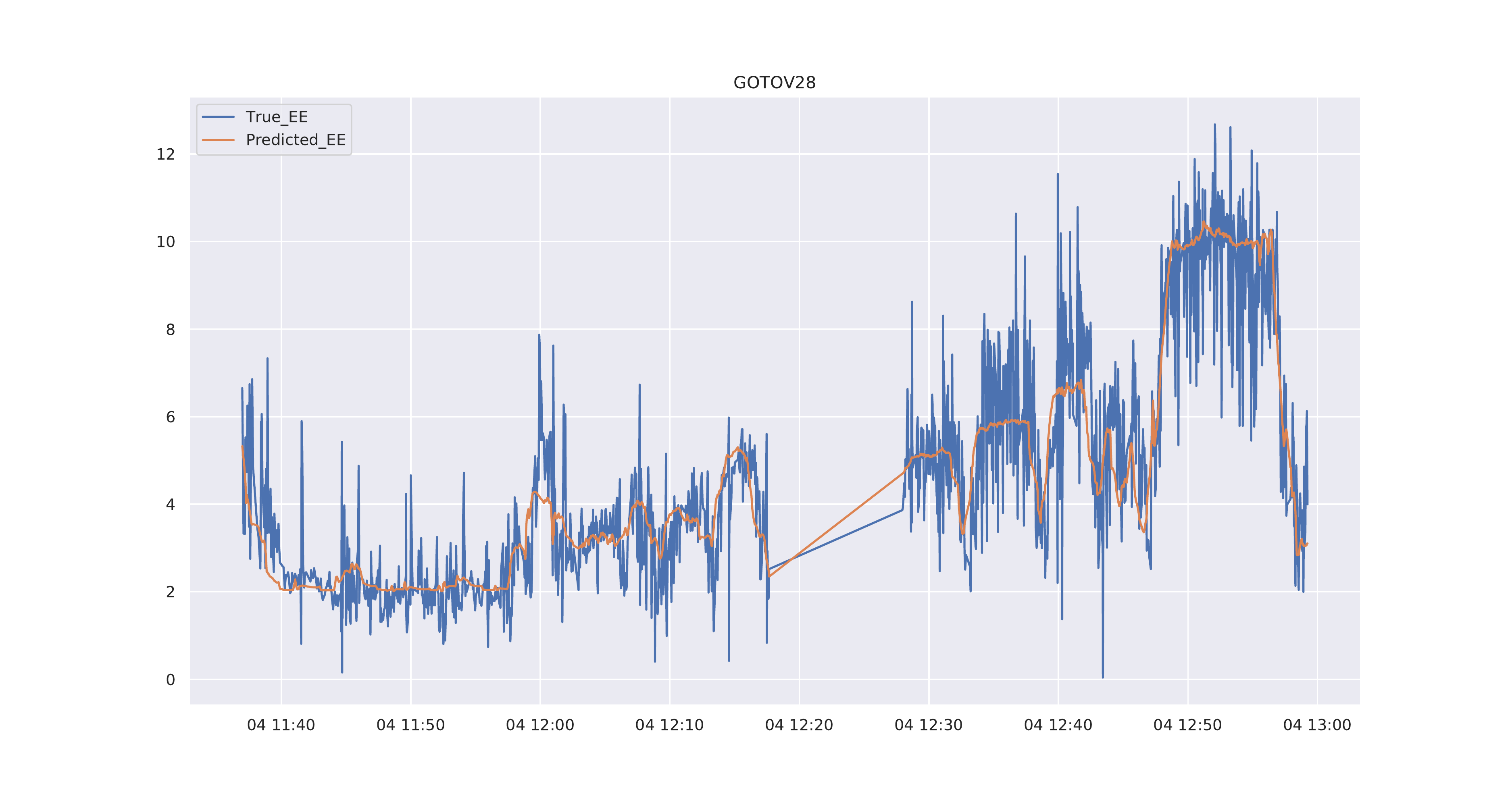}
    \caption{True versus Predicted EEm, one participant example.}
    \label{fig:true_pred_EEm_onePart}
\end{figure}

\begin{figure}[t]
    \centering
    \begin{minipage}{.5\textwidth}
        \centering
        \includegraphics[clip, trim= 1cm 1cm 1.8cm 2cm, width=\textwidth]{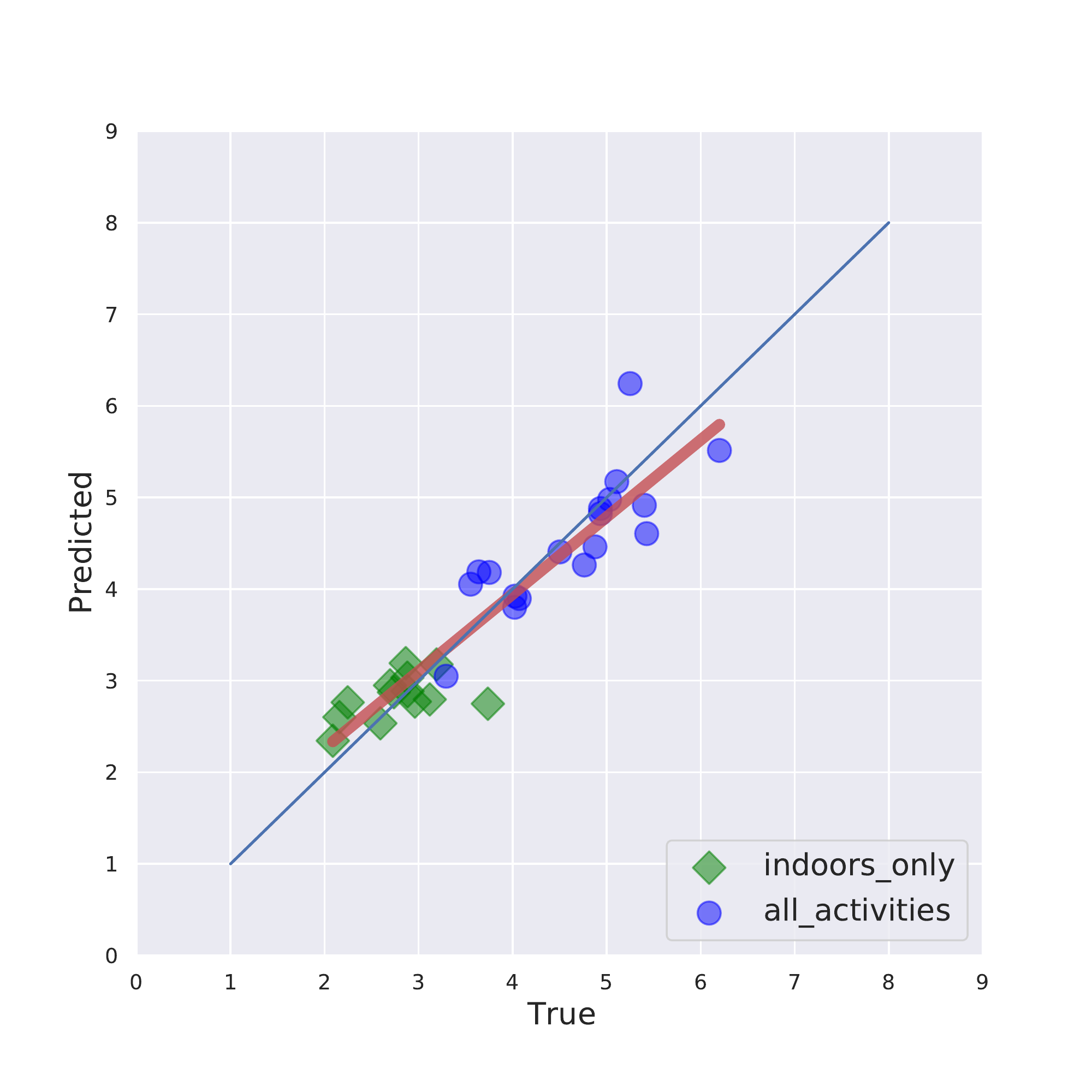}
    \end{minipage}%
    \begin{minipage}{0.5\textwidth}
        \centering
        \includegraphics[clip, trim= 1cm 1cm 1.8cm 2cm, width=\textwidth]{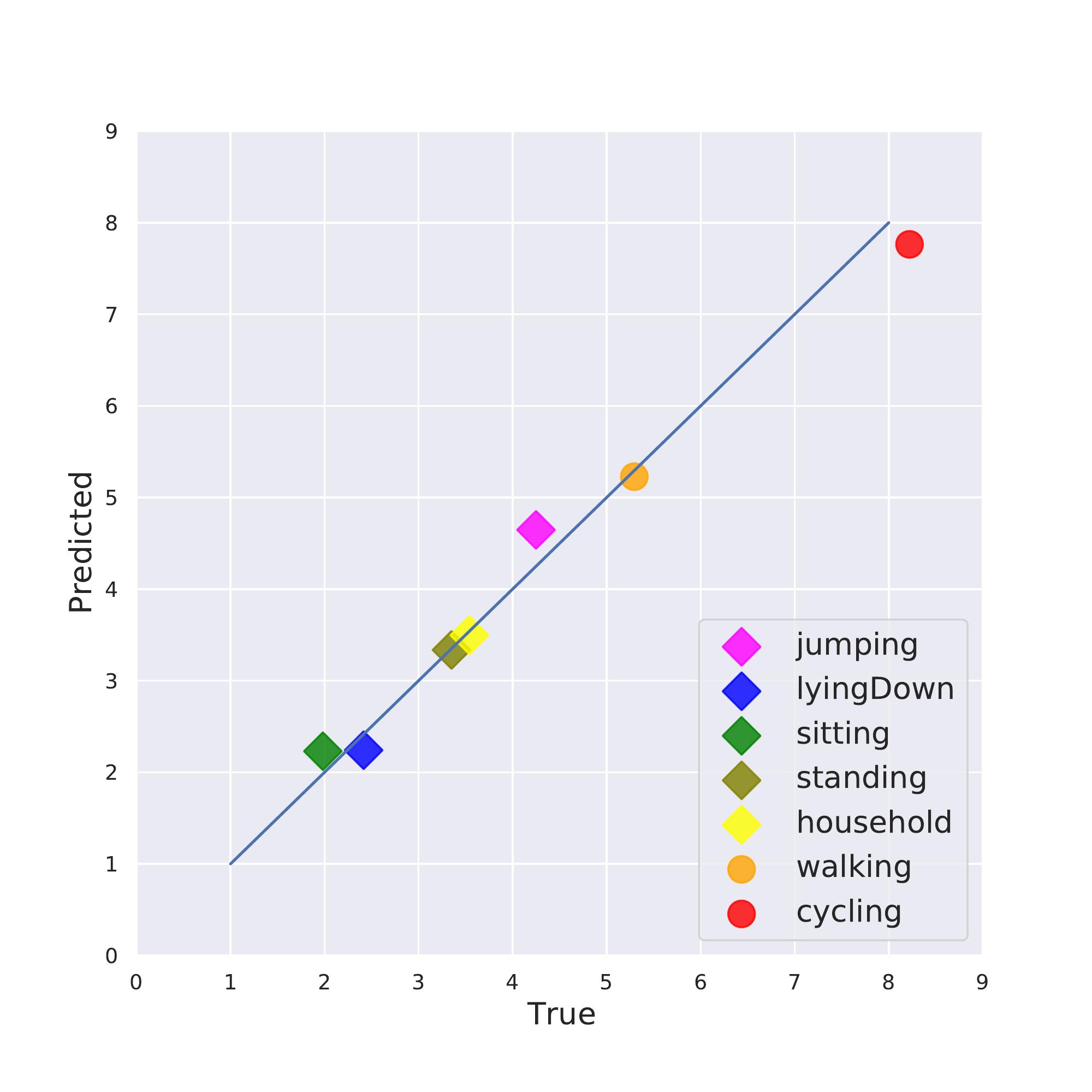}
    \end{minipage}
    \caption{Scatter plot of mean true over mean predicted EEm, per participant (left) and activity (right).}
    \label{fig:true_pred_EEm}
\end{figure}

Concluding, the model that combines accelerometry with individuals' details (GA\_ID) result to significantly ($p<0.05$) higher R$^2$ and lower RMSE when compared to the only accelerometer data (GA). Activity knowledge, however, does not add any extra value to the PAEE estimation challenge. To demonstrate the general performance of the RNN model with SD as aggregation function and besides accelerometer data also individual level data as input, we plotted Figures~\ref{fig:true_pred_EEm_onePart} and \ref{fig:true_pred_EEm}.

In Figure~\ref{fig:true_pred_EEm_onePart}, we plotted the predicted over true PAEE (COSMED) values for one participant. Here, we see that the model overall predicts really close to the mean of true EEm, as the orange line (predicted EEm) very nicely follows the trend of the blue line (true EEm). The gap in the middle of the plot is the transition from indoor to outdoor activities, where there were no COSMED measurements.

In more detail, in Figure~\ref{fig:true_pred_EEm}, we see two scatter plots of the average true over average predicted EEm value per participant (left) and per activity class (right). In the left plot, the blue dots display the participants with both indoor and outdoor activities, while the green diamonds represent participants with only indoors data. Adding to that, the red trend line is used to compare the predictions to the main diagonal (blue, representing $x=y$) which is the ground truth. From that, we observe that the model has on average good performance. However, it slightly overestimates the lowest EEm values (red line above main diagonal) and underestimates the highest ones (red line below main diagonal). 

In the right plot of Figure~\ref{fig:true_pred_EEm}, we can see that our model captures really well the average EEm per activity since the average PAEE estimated for all the activities (colored dots) are almost either on or really close to the ground truth blue line. However, evaluating the performance over one activity class averages out some of the over- and underestimations.

Finally, in Table~\ref{tab:resultsBest}, we present the performance of the GA\_ID model by different target aggregations. We aggregated the original and predicted EEm values at $10, 30$ and $60$ seconds, and at $5$ and $60$ minutes. This is really useful, since in a free-living setting, the model will be used to estimate PAEE for aggregated window. From the table, it is observed that even with the shorter window of aggregation, $10$ seconds, the model's performance improves substantially both for R$^2$, from $0.55$ to $0.65$, and RMSE, from $1.25$ to $0.95$ Kcal/min. Additionally, for the $60$ seconds window, a time frame that is commonly used in the literature~\cite{Staudenmayer2009, Ellis2014, Driscoll2020}, the model has an RMSE of only $0.76$ Kcal/min with the predictions explaining  $78\%$ of the original EEm signal variation.

\begin{table}[t]
\centering
\caption{Comparing true and predicted EEm over different aggregations.}
\label{tab:resultsBest}
\begin{tabular}{@{}l|
>{\columncolor[HTML]{EFEFEF}}c c
>{\columncolor[HTML]{EFEFEF}}c |c
>{\columncolor[HTML]{EFEFEF}}c c@{}}
\toprule
\textbf{Aggregation} & {\color[HTML]{000000} \textbf{R$^2$}} & {\color[HTML]{000000} \textbf{inR$^2$}} & {\color[HTML]{000000} \textbf{outR$^2$}} & \textbf{RMSE} & \textbf{inRMSE} & \textbf{outRMSE}  \\ \midrule
\textbf{per breath}  & $0.55$                                & $0.37$                                  & $0.36$                                   & $1.25$        & $1.09$          & $2.04$             \\
\textbf{10 sec}      & $0.65$                                & $0.48$                                  & $0.50$                                   & $0.95$        & $0.89$          & $1.58$             \\
\textbf{30 sec}      & $0.72$                                & $0.58$                                  & $0.56$                                   & $0.82$        & $0.74$          & $1.40$             \\
\textbf{60 sec}      & $0.78$                                & $0.66$                                  & $0.62$                                   & $0.76$        & $0.64$          & $1.34$             \\
\textbf{5 min}       & $0.82$                                & $0.74$                                  & $0.63$                                   & $0.62$        & $0.48$          & $0.97$             \\
\textbf{60 min}      & $0.81$                                & $0.63$                                  & $0.49$                                   & $0.40$        & $0.30$          & $0.77$              \\ 
\bottomrule
\end{tabular}
\end{table}

\section{Discussion}
\label{sec:conclusion}
In this paper, we developed and tested a recurrent neural network to estimate older adults' physical activity energy expenditure. The model architecture combining an RNN with $3$ GRU layers with a feedforward network of one dense layer seem to cope well with sequential data like accelerometer time-series without the need of any sophisticated feature construction step. Remarkably, when GRU layers are feed with accelerometer data aggregated with statistical dispersion metric such as SD or IQR instead of averaging by mean, not only the model's performance is significantly improved but this improvement is achieved by using sequences with approximately $10$ times smaller size. In addition, taking besides the accelerometer data, participants demographic and body composition information into account, our model performance improved with up to $10\%$. The explained variance of the optimal RNN is $0.55$ when evaluated per breath and can go up to $0.80$ when computed per minute.

For the estimation of PAEE using the developed RNN, no sophisticated feature construction step is required. In fact, accelerometer data aggregated with SD to a sampling rate of $0.42$ Hz with participant-level and tested per minute EEm aggregation windows, can explain most of EEm signal variation. This is because statistical dispersion metrics can represent the original signal in a more characteristic way compared to averaging with mean~\cite{Lyden2011}. Furthermore, for the estimation of PAEE using the developed RNN, a two-minutes window size can be represented with only $50$ inputs per sequence without significant loss of accuracy when compared to four minute window with $480$ inputs (mean). The developed RNN is a computationally efficient, fast and accurate method to estimate PAEE in older adults.

The PAEE estimations improve as expected when participant-level data are taken into account. In detail, when combining demographic and anthropometric data with accelerometry (model GA\_ID) and testing them on the original COSMED output (per breath) the RMSE is reduced from $1.35$ to $1.25$ Kcal/min and R$^2$ increased from $0.45$ to $0.55$, with the improvement mainly showing in the lower intensity activities (indoors) where RMSE decreased from $1.16$ to $1.09$ Kcal/min. However, the addition of activity classes, even when they are combined with participant-level and accelerometer data (model GA\_ID\_AC), did not give any significant improvement to the models. This is an important observation for our architecture since it comes in contrast with what is shown from previous work of Altini~\cite{Altini2015}. It seems that the way RNNs model the input sequences and its ability to ``remember'' past information, the exact activity labels are not needed for efficient PAEE estimations. Therefore, when the objective is only PAEE estimation and not its association with specific activities, there is no need of applying activity recognition algorithms beforehand.

\begin{figure}[t]
    \centering
    \includegraphics[clip, trim= 2.5cm 1cm 4cm 1cm, width=\textwidth]{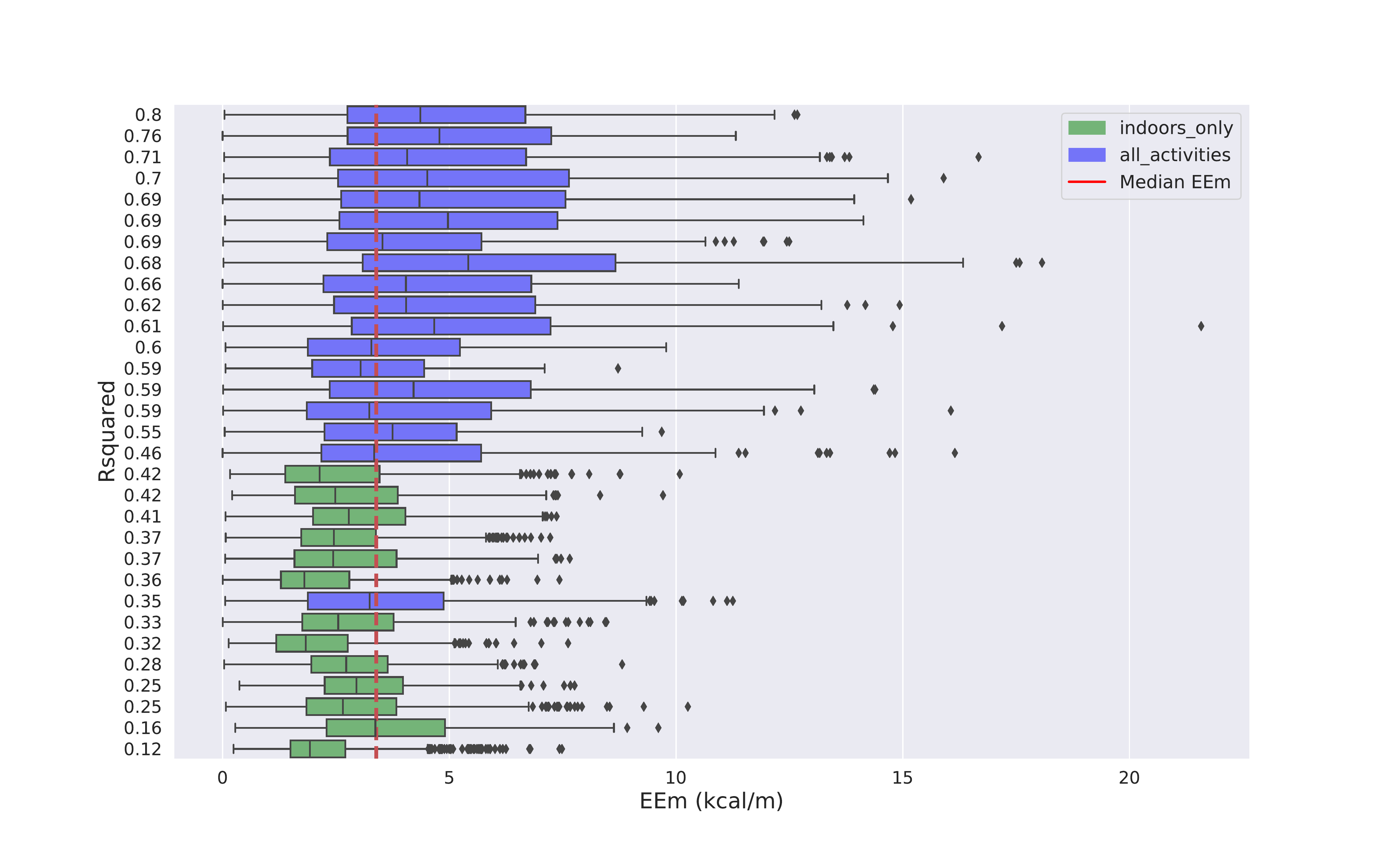}
    \caption{Box plots of mean True EEm per participant ordered by $R^2$.}
    \label{fig:boxplot_EEm_perPart}
\end{figure}

We observe that the model-fit is somewhat different for indoor and outdoor activities. The RMSE is equal or less than $1.00$ Kcal/min for indoor activities, when comparing them for windows of $30$ and $60$ seconds, while for outdoors, it is $1.20$ for walking and $1.50$ for cycling. But when we compare our model performance per participant, we see a slight overestimation of average EEm for those with only indoor activities (low intensity activities) and a slight underestimation for those with high average EEm. Especially when ordering by R$^2$ (see Figure~\ref{fig:boxplot_EEm_perPart}), we can clearly see that for participants with lower median EEm, the model explains less of its variation (lower R$^2$ ) while for participants with longer range of EEm values, the model captures most of their EEm variation. However, that is also due to the fact that for participants with smaller EEm values, even smaller RMSE errors can really lead to lower R$^2$. This can be seen from the result tables where, indoors RMSE are lower compared to outdoors, but they produce lower R$^2$ values. We conclude that the model on average performs well both for high and low intensity activities since their averaged predicted values are really similar to the true ones (see Figure~\ref{fig:true_pred_EEm}).

During the development of our models, we realized that the GOTOV dataset involves some data collections limitations. Indirect calorimetry was collected in a continuous way with small in between activity breaks (max $1$ minute) to recognize the separate activities. These rather small breaks between activities make it difficult to estimate the EEm outcome of specific activities due to the energy expenditure’s lag effect. In detail, without long discriminated breaks between activities, it is highly possible that past activities influence the EEm records of future ones. Fortunately, due to the RNN modeling advantage, we managed to take in account preceding activity information by incorporating data of longer windows ($2$ to $4$ minutes) and letting the model decide on which information from the window is the one with higher weight to EEm estimation. The great advantage of the GOTOV dataset is that there are a satisfactory number of participants and that this data set is dedicated to people over $60$ years of age. Because PAEE monitoring in elderly has large potential to stimulate vital and healthy ageing, the GOTOV dataset is perfectly suited for the development of activity recognition and PAEE estimation models.

In summary, the results of this study demonstrate that RNNs can be a solution to the challenge of PAEE estimation. While they do not require any complex feature construction steps and can be trained with way lower resolution accelerometer data, if this is resampled with statistical dispersion metrics, RNNs produce PAEE estimations similar or better than competing methods. Because RNNs take into account longer windows in activity history without increasing the size and dimensionality of their input information, we believe that such a modeling technique can be proven really advantageous when it is applied to free-living accelerometer data that is collected in a continuous way.

Applying such a model to free-living data collections was one of the motivations of our study. In our future work, we intent to apply our modeling technique to free-living intervention studies on older individuals. From such an application, we expect to have a deeper understanding of how changes in PAEE levels and activity types could influence the health of older individuals and potential further stimulate vital and healthy ageing. In order to achieve this, we aim to build characteristic features of PAEE levels and PA types of long time periods (weeks, months) and relate them with parameters of metabolic health, general health and well-being. These relations, then, can be turned into distinct recommendations for effectively maintaining mobility among older adults and a continuous monitoring system to track the adherence and improvement of metabolic health.

\newpage


\bibliographystyle{unsrt}  
\bibliography{main.bib} 

\end{document}